\RequirePackage[2020-02-02]{latexrelease}
\documentclass[aps,a4paper,superscriptaddress,nofootinbib,12pt]{revtex4}
\usepackage[T1]{fontenc}
\usepackage{graphicx}
\usepackage{hyperref}
\usepackage{amsmath}
\usepackage{amsthm}
\usepackage{amsfonts}
\usepackage{xcolor}
\usepackage{amssymb}
\usepackage{stmaryrd}
\usepackage[toc,page]{appendix}

\begin{document}

\title{Non-homogeneous exoplanets in metric-affine gravity}

\author{\footnotesize ALEKSANDER KOZAK}

\address{Institute of Theoretical Physics, University of Wrocław, pl. Maxa Borna 9,\\
50-204 Wrocław, Poland\\
aleksander.kozak@uwr.edu.pl
}

\author{\footnotesize ANETA WOJNAR}

\address{Laboratory of Theoretical Physics, Institute of Physics, University of Tartu, W. Ostwaldi 1,\\
50411 Tartu, Estonia\\
aneta.magdalena.wojnar@ut.ee
}
\begin{abstract}
The improved description of the planets' interior is provided. We examine the modified gravity effects on the Earth-like planets composed of the iron core and silicate mantle. We confirm that the mass-radius relations as well as density profiles differ with respect to the commonly adopted Newtonian models.
\keywords{(exo)-planets, modified gravity.}
\end{abstract}

\maketitle

\bibliographystyle{wsp}

\bibliography{\jobname}

\section{Introduction}

General Relativity (GR) is one of the most thoroughly tested physical theories \cite{Will:2014kxa}. It was not only able to explain various phenomena that could not be accounted for on the ground of previous theory, but also successfully predicted existence of some exotic objects, such as black holes. Evidence for their existence was provided by a recent detection of gravitational waves \cite{TheLIGOScientific:2017qsa}, as well as a direct observation of a shadow of a black hole at the centre of M87 \cite{Akiyama:2019cqa,aki2,aki3,god}. However, not all cosmological and astrophysical data fits into the scenario predicted by GR. For example, it is unclear how the currently observed accelerated expansion of the Universe should be explained, as well as the problem of dark matter. These problems were addressed by modified gravity theories \cite{Copeland:2006wr,Nojiri:2006ri,nojiri2,nojiri3,Capozziello:2007ec,Carroll:2004de}, introducing some deviations from Einstein's theory. Within the context of modified gravity, the space-time singularities and high-energy unification scenarios were also investigated \cite{Senovilla:2014gza,ParTom,BirDav}. Another major problem that cannot be satisfactorily solved by GR is the excess of maximum mass of compact objects and binary black hole mergers as compared to theoretical predictions \cite{lina,as,craw,NSBH,abotHBH,sak3}. 

In this paper, we focus on a narrow class of modified theories of gravity, called $f(R)$ theories, in order to investigate how their non-relativistic limit affects internal composition of low-mass objects. The problems sketched above concern mostly high-energy regime or large, cosmological scales, but it might be possible that also weak-field limit will be altered, allowing one to detect deviation from GR on much smaller scales. For example, it has been shown that non-relativistic description of objects such as main-sequence and white dwarf stars is changed in modified gravity \cite{Saito:2015fza,olek,olmo_ricci,review,cantata,Chandra,Saltas:2018mxc,Jain:2015edg,Banerjee:2017uwz,Wojnar:2020wd,Belfaqih:2021jvu,Sakstein:2015zoa,Sakstein:2015aac,Crisostomi:2019yfo,Olmo:2019qsj,Rosyadi:2019hdb,early,chow,Benito2021,Wojnar:2020frr}.  
On the other hand, planets in the context of alternative theories of gravity were not studied too much in the literature. The evolution of Jupiter-like planets and fragmentation process were investigated in \cite{an_jowisz}, while mass-radius relation in \cite{olek2}. Apart from these works, they have been mainly used to test and constrain theories of gravity in the case of precessions of planetary perihelion in the Solar System, see e.g. \cite{BD,melinda,iorio,harko,hatice,bag,seba,schmidt,harko2,iorio2,fatibene,bhat}, as well as modifications to the third Kepler's law \cite{kepler1,kepler2}.

We adopt so-called 'Palatini formalism', in which one considers the metric tensor to be a priori independent of the affine connection. This is motivated by the fact that logically these two structures - affine and metric, the former being responsible for parallel transport and the latter for measurement of distances, time, volumes, and also for setting up causal structure of spacetime - are unrelated. The connection is assumed to be symmetric, it also does not enter the matter part of the action. Thanks to this assumptions, the theory is metric, i.e. particles follow geodesics determined by the dynamical metric. The Palatini formalism, when applied to $f(R)$ theories, exhibits some advantages over the metric approach. For example, one does not need any screening mechanism to hide the effect of a scalar degree of freedom present in the metric theory, since the field is not dynamical \cite{olmo2011}. Also, they provide a framework in which the inflation appears in a natural way \cite{borowiec2016}. 

In the previous work \cite{olek2} we have revealed that in the mass-radius relations for a cold homogeneous sphere one deals with the additional degeneracy caused by metric-affine gravity, and that particular models of gravity modify the interior structure of terrestrial planets. Since that finding carries a possibility to test theories of gravity with the use of earth- and marsquakes, as detailed discussed in \cite{olek2}, it is crucial to improve our planets' modelling by taking into account a more complex interior structure and realistic geometry. Therefore, in what follows, we will improve our analysis by adding a second layer - our toy planet will now consist of an iron core and silicate mantle, in the proportions adequate to the description of the Earth-like exoplanets. Before doing so, in the next section we will briefly recall the theoretical background and structural equations with well-tested equations of state for the iron and silicates. In the section (\ref{num}) we present the boundary conditions and numerical solutions of the modified equations in the form of the mass-radius relations and density profiles for the Palatini quadratic model. The last section is dedicated to the short conclusions. 

We use $(-+++)$ signature convention and $\kappa^2=8\pi G/c^4$.

\section{Theoretical background}
\subsection{{Conformal invariants approach to Palatini $f(\mathcal R)$ gravity}}
General action for metric scalar-tensor theories of gravity reads as follows:
\begin{equation}\label{generalAction}
    \begin{split}
        S[g,\Phi, \psi_m]=&\frac{1}{2\kappa^2}\int \sqrt{-g}\Big[\mathcal{A}(\Phi) R(g) -\mathcal{B}(\Phi)(\partial\Phi)^2\\
       & - \mathcal{V}(\Phi)\Big]d^4 x
+S_{\text{matter}}[e^{2\alpha(\Phi)}g,\psi_m].
    \end{split}
\end{equation}
with $\psi_m$ denoting matter fields. There are four functions of $\Phi$ entering the action: $\mathcal{A}, \mathcal{B}$ and $\alpha$ denoting different types of coupling: to the curvature, to the kinetic term and anomalous to matter, respectively, and also the self-interaction potential of the scalar field, represented by the function $\mathcal{V}$. This action will provide mathematical framework for analysis of Palatini $f(\mathcal{R})$ gravity given by the action:
\begin{equation}
S[g,\Gamma,\psi_m]=\frac{1}{2\kappa^2}\int \sqrt{-g}f(\mathcal{R}) d^4 x+S_{\text{matter}}[g,\psi_m],\label{action}
\end{equation}
where $\mathcal{R} = g^{\mu\nu}\mathcal{R}_{\mu\nu}(\Gamma)$\footnote{In what follows we will denote the Palatini curvature by a calligraphic $\mathcal{R}$, and the metric curvature by $R$.}. Here, $\Gamma$ is treated as an object independent of the metric tensor and scalar field. 

The theory
has a scalar-tensor representation in the metric formulation:
\begin{equation}\label{palatiniSTAction}
    \begin{split}
        S[g,\Phi,\psi_m]=&\frac{1}{2\kappa^2}\int \sqrt{-g}\Big[\Phi R(g) +\frac{3}{2\Phi}(\partial\Phi)^2\\
       & - \mathcal{V}(\Phi)\Big]d^4 x
+S_{\text{matter}}[g,\psi_m],
    \end{split}
\end{equation}
that can be achieved by means of a Legendre transformation: $\Phi = df/d\mathcal{R}$ and $V(\Phi) = f'(\mathcal{R}(\Phi))\mathcal{R}(\Phi) - f(\mathcal{R}(\Phi))$ and integrating out the auxiliary connection field.
The four functions of the scalar field from \eqref{generalAction} can be now identified with: 
\begin{equation}
\begin{split}
    & \mathcal{A}(\Phi) = \Phi, \quad \mathcal{B}(\Phi) = -\frac{3}{2\Phi}, \\
    & \mathcal{V}(\Phi) = V(\Phi), \quad \alpha(\Phi) = 0.
\end{split}
\end{equation}
Variation w.r.t. dynamical variables entering the action \eqref{palatiniSTAction} leads to the following field equations:
\begin{subequations}\label{fieldEquations}
\begin{align}
\begin{split}
& \Phi G_{\mu\nu}-\frac{3}{4\Phi}g_{\mu\nu}g^{\alpha\beta}\partial_\alpha\Phi\partial_\beta\Phi +\frac{3}{2\Phi}\partial_\mu\Phi\partial_\nu\Phi \\
&  - (g_{\mu\nu}\Box - \nabla_\mu\nabla_\nu)\Phi +\frac{1}{2}V(\Phi) g_{\mu\nu} = \kappa^2 T_{\mu\nu},
\end{split} \\
& V(\Phi)-\frac{1}{2}\Phi V'(\Phi)=\frac{\kappa^2}{2} T\,.
\label{scFiEOM}
\end{align}
\end{subequations}
One can introduce Weyl transformations of the metric tensor in order to change the parametrization of \eqref{generalAction} in such a way that the field equations involving new functions $\{\mathcal{A}, \mathcal{B}, \mathcal{V}, \alpha\}$ will be more approachable. The transformation is defined as:
\begin{eqnarray}\label{transformation}
\begin{cases}
\bar{g}_{\mu\nu} = e^{2\gamma(\Phi)}g_{\mu\nu} \\
\bar{\Phi} = \bar{f}(\Phi),
\end{cases}
\end{eqnarray}
where one allows the scalar field to change as well. Under these transformations, the four functions defining so-called conformal frame also change. It is possible, however, to construct quantities preserving their functional form when \eqref{transformation} is applied to dynamical variables. Such quantities will be called (conformal) invariants; the ones used in this paper are the following:
\begin{eqnarray}
\mathcal{I}_1(\Phi)& =& \frac{\mathcal{A}(\Phi)}{e^{2\alpha(\Phi)}}, \\
\mathcal{I}_2(\Phi)& =& \frac{\mathcal{V}(\Phi)}{\mathcal{A}^2(\Phi)},\\
\frac{d\mathcal{I}}{d\Phi}& =& \sqrt{\frac{\mathcal{B}}{\mathcal{A}}+ \left(\frac{3\mathcal{A}'}{2\mathcal{A}}\right)^2},
\end{eqnarray}
where prime denotes differentiating w.r.t. the scalar field. For Palatini $f(R)$ theory, the invariants are:
\begin{eqnarray}
\mathcal{I}_1(\Phi)& =& \Phi, \\
\mathcal{I}_2(\Phi)& =& \frac{V(\Phi)}{\Phi^2},\\
\frac{d\mathcal{I}}{d\Phi}& =& 0,
\end{eqnarray}
with the potential function left unspecified (it is determined by the choice of the $f$ function). Each of the invariants introduced above has a meaning within the context of scalar-tensor theories of gravity. The first one represents non-minimal coupling between the scalar field and curvature, and if it is constant then it will be always possible to find a conformal frame, in which $\Phi$ can be viewed as just another matter field. $\mathcal{I}_2$ generalizes the self-interaction potential, while $\mathcal{I}$ plays the role of invariant scalar field. Its vanishing means that the field is non-dynamical for a certain class of theories; conformal transformations do not change this fundamental property of such theories. As one can see, the invariant vanishes for Palatini $f(\mathcal{R})$ gravity, which is a well-known fact. Since the field is non-dynamical for this theory, it can be related in an algebraic way to the trace of stress-energy tensor, as suggested by the field equations \eqref{fieldEquations}.

\subsection{Planets' structure equations in Palatini gravity}

The hydrostatic equilibrium equations for the Palatini $f(\mathcal{R})$ gravity in the case of the quadratic functional 
$$
f(\mathcal R)=\mathcal R + \beta \mathcal R^2
$$
and the perfect fluid energy-momentum tensor $T_{\mu\nu}=(\rho+p)u_\mu u_\nu +pg_{\mu\nu}$
were presented in various works (see \cite{aneta_stab,olek}). In what follows, we will need their simplified forms since the terms proportional to $p/c^2$ can be neglected in comparison to the energy density. Therefore, the structural equations are \cite{olek2}
\begin{equation}\label{tovJ1}
    \begin{split}
        & p' =  -\frac{G\mathcal{M} \rho}{r^2\mathcal{I}^{1/2}_1}\left(1 - \frac{2G\mathcal{M}}{c^2r\mathcal{I}^{1/2}_1}\right)^{-1}\left(1 + \frac{4\pi\mathcal{I}_1^\frac{3}{2}r^3}{c^2\mathcal{M}}\frac{\mathcal{I}_2}{2\kappa^2}\right)\left(\frac{r}{2}\partial_{r} \ln\mathcal{I}_1 + 1 \right) \\
        & 
   -c^2\rho \:\partial_{r} \ln\mathcal{I}_1
    \end{split}
\end{equation}
with the invariants taking the following forms for the given model of gravity: 
\begin{align*}
    & \mathcal{I}_1 = 1 + 2\alpha\rho ,\;\;\;\;\mathcal{I}_2 = \frac{2\alpha\kappa^2c^2\rho^2}{\left(1 + 2\alpha\rho  \right)^2}.
\end{align*}
We have introduced a redefined Starobisnki parameter
$  \alpha := 2 c^2 \kappa^2 \beta$. Moreover, the
non-relativistic mass was reduced to:
\begin{equation}\label{mass1}
    \mathcal{M}(r) =    \int^{r}_0 4\pi\tilde{r}^2 \frac{\rho - \alpha\rho^2}{\left(1 + 2\alpha\rho  \right)^{1/2}}
   \left[1+ \frac{\tilde{r}}{2}\partial_{\tilde{r}}\ln\left(1+2\alpha\rho \right)\right]d\tilde{r}.
\end{equation}
Apart from the structural equations, one needs to have an equation of state (EoS) which describes the iron core and silicate mantle in our 2-layer model of a terrestrial planet, and boundary conditions, discussed in the section (\ref{num}). Since we are also dealing with the pressure ranges $p\gtrsim10^4\,\text{GPa}$, the electron degeneracy must be taken into account. To do so, one matches the Birch EoS \cite{birch,poi} 
with the Thomas-Fermi-Dirac EoS (TFD) \cite{thomas, fermi, dirac, feynmann}. The last one also takes into account a density-dependent correlation energy terms \cite{sal} being a result of interactions between electrons themselves which obey the Pauli exclusion principle and move in the Coulomb field of the nuclei. Such a merger can be approximated by a modified polytropic equation of state (SKHM EoS) \cite{seager}
\begin{equation}\label{pol}
    \rho(p)=\rho_0 +cp^n,
\end{equation}
whose best-fit parameters $\rho_0$, $c$, and $n$ for iron and silicate (Mg, Fe)SiO$_3$ are given in the table \ref{tabpoly}. The additional term $\rho_0$ includes the incompressibility of solids and liquids at low pressures. The validity of the equation (\ref{pol}) with the given fits reaches  $p=10^{7}\,\text{GPa}$, and hence it will be the maximal value of the central pressure considered in our numerical analysis.

\begin{table}
\caption{Best-fit parameters for the SKHM EoS (\ref{pol}) obtained in the reference\cite{seager}.}
\centering
\begin{tabular}{llll }
\hline\noalign{\smallskip}
Material & $\rho_0$ (\text{kg m}$^{-3}$) & $c$ (\text{kg m}$^{-3}$ \text{Pa}$^{-n})$ & $n$ \\
\noalign{\smallskip}\hline\noalign{\smallskip}
Fe($\alpha$) & 8300 & 0.00349 & 0.528\\
 (Mg, Fe)SiO$_3$ & 4260 & 0.00127 & 0.549  \\
\noalign{\smallskip}\hline
\end{tabular}\label{tabpoly}
\end{table}

\section{Numerical solutions}\label{num}
Since we are interested in the non-homogeneous planets, 
we need to use separate equations of state for different layers. In this paper, we limit our attention to spherical objects composed of two layers: the inner iron core and the mantle made of silicate perovskite (Mg, Fe)SiO$_3$. Such a chemical composition was chosen to render more realistically structure of rocky planets such as Earth or Earth-like exoplanets.

{Before going further, let us briefly discuss matching the core and the mantle. Junction conditions in the case of Palatini were studied in \cite{olmojunction} - let us however notice that our case differs from it, since in that work interior solution was matched with the exterior one, providing additional conditions on the matter fields. In our case, our main concern is related rather to the so-called "singular parameter", discussed in \cite{olek2}, since it was evident that for some particular combination of the parameter's value and equation of state the hydrostatic equilibrium equation could not have any physical solutions. Therefore, one needs to be careful when performing the calculations, especially for negative values of the parameter $\alpha$.}

In order to obtain density profiles of planets, one needs to make use of Eqs \eqref{tovJ1}, \eqref{mass1} and \eqref{pol} to write down the considered system of equations in the following form:
\begin{eqnarray}
\begin{cases}
\rho'(r) = f_1(\rho, \mathcal{M}, r), \\
\mathcal{M}'(r) =  f_2(\rho, \mathcal{M}, r),
\end{cases}
\end{eqnarray}
with the {boundary} conditions $\mathcal{M}(0) = 0$ and $\rho(0) = \rho_c$, integrated until the density drops to the ambient one. If one wishes to find the density profile for a planet of a given mass $m$ (i.e. solve the above system for $\mathcal{M}(0) = 0$ and $\mathcal{M}(R_p) = m$, where $R_p$ is the planet radius), then it is necessary to find the internal density first; in our case, it was achieved by a simple shooting method. We also set the core mass fraction to be 0.3, based on the most common Earth and Earth-like models of terrestrial planets \cite{icarus}. At the core-mantle boundary we demand that the core pressure and the mantle one are equal, which gives a condition for the initial density in the second layer. The numerical calculations were performed using a Python script.

\begin{figure}[h]
\centering
    \includegraphics[width=1\linewidth]{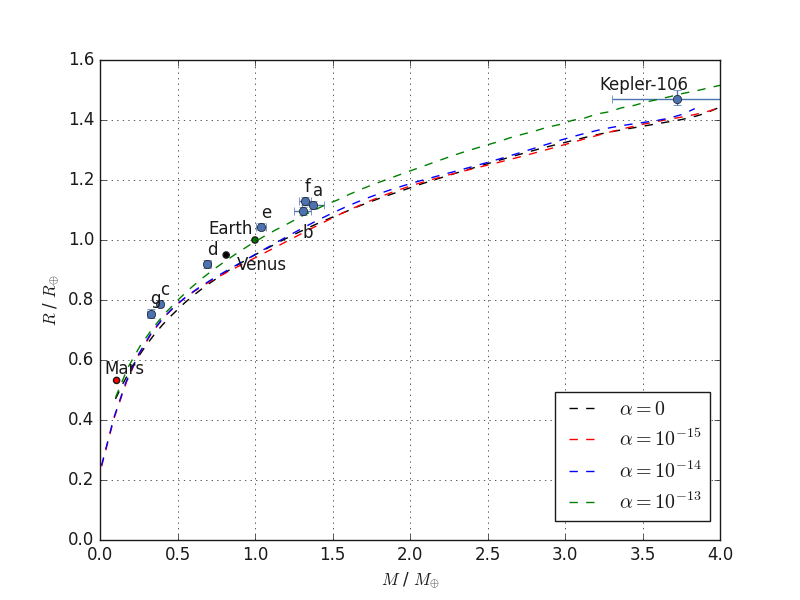}
\caption{[color online] Mass-radius relation for small planets composed of two layers: iron core and perovskite silicate mantle. The results were obtained for different values of the Starobinsky parameter $\beta$. The solar-system planets were included, as well as some TRAPPIST-1 exoplanets, denoted by letters \cite{trappist}.}
\label{fig1a}
\end{figure}

\begin{figure}[h]
\centering
    \includegraphics[width=1\linewidth]{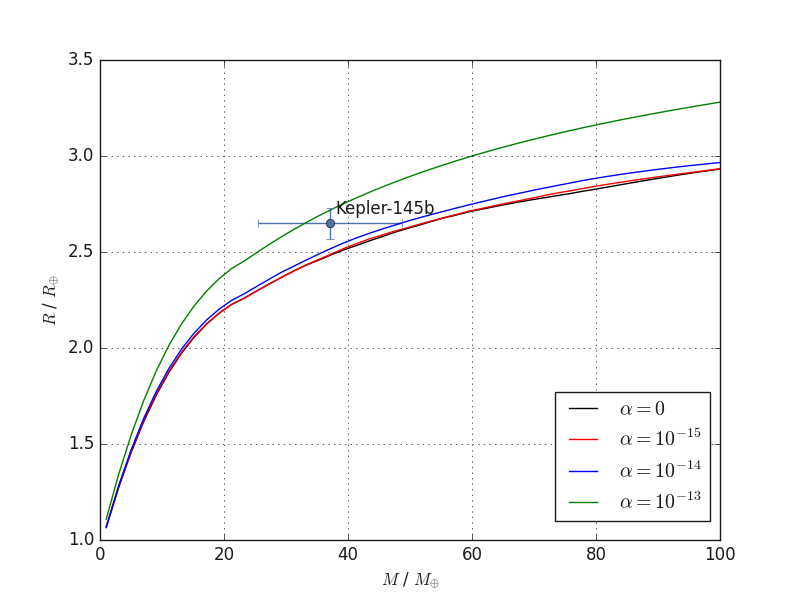}
\caption{[color online] Mass-radius relation for large planets composed of two layers: iron core and perovskite silicate mantle. The results were obtained for different values of the Starobinsky parameter $\beta$. We added a point representing mass and radius of Kepler-145b exoplanet \cite{xie} to illustrate that the effect of modified gravity can be large for bigger planets.}
\label{fig1b}
\end{figure}

We present the results of numerical integration in the figures above and below. Figures \ref{fig1a} and \ref{fig1b} show the mass-radius relation for small and bigger planets, respectively, for various values of the Starobinsky parameter $\beta$. Here, we use the parameter $\alpha = 2c^2\kappa^2\beta$ to quantify deviation from GR. We see that the bigger values of the parameter correspond to bigger (and less dense) planets of the same mass. This can be also seen when density profile, shown in Figure \ref{fig2}, is investigated. Here, we plotted the profiles for planets of mass equal to the Earth's one. It means that, for different values of $\beta$, they will have various values of the central density. It can be seen that there is a tendency toward smaller densities (and hence pressures) in the interior of planets for positive values of $\beta$. Also, it might seem surprising that the mass is the same in all three cases, since the area under the curve corresponding to larger value of $\beta$ is smaller than the other three. However, we must remember that the mass is not simply obtained by integrating $\int_0^{R_p} 4\pi r^2\rho dr$, but there are also contributions from modifications of the theory (see the equation \eqref{mass1} and detailed discussion in \cite{olek2} and \cite{delacruz}). One must notice that in the Palatini gravity, the effects coming from modifications of geometry do not contribute to the overall mass outside the sphere, as the density of matter vanishes there and hence, the integrand in \eqref{mass1}. 

\begin{figure}[h]
\centering
    \includegraphics[width=1\linewidth]{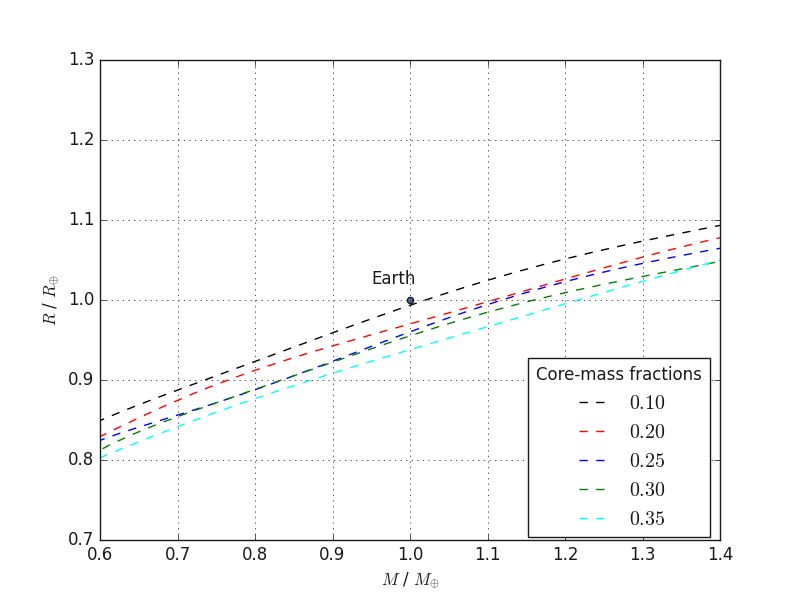}
\caption{[color online] {Mass-radius relation for small planets with different core-mass fractions when there is no modification of gravity taken into account. The blue dot represents planet Earth. The calculations were carried out assuming two-layer composition of the planets (iron core and perovskite silicate mantle).}}
\label{fig3}
\end{figure}

\begin{figure}[h]
\centering
    \includegraphics[width=1\linewidth]{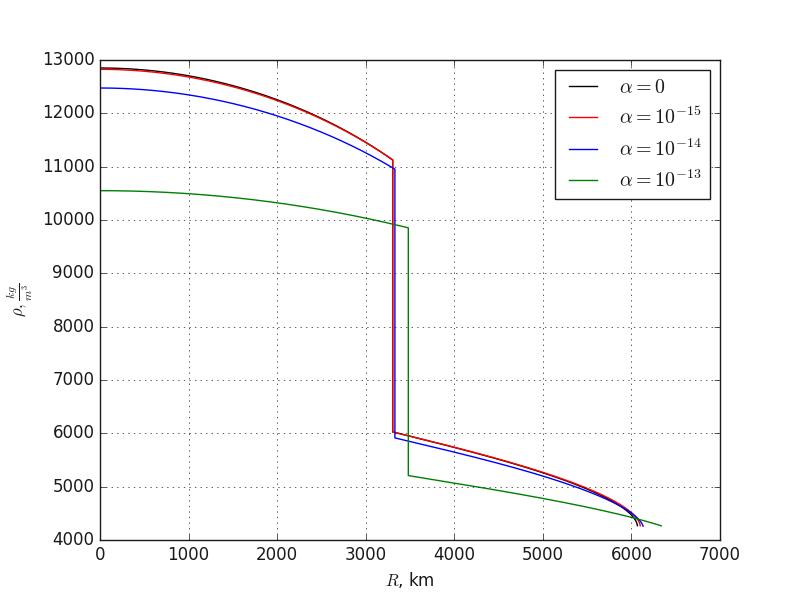}
\caption{[color online] Density profile for planets composed of two layers. Different curves correspond to different values of the Starobinsky parameter $\beta$.}
\label{fig2}
\end{figure}

{One can also ask why the best-fit curve for the planet Earth in the Figure \ref{fig1a} is the one for $\alpha = 10^{-13}$, which seems to be in conflict with well-established data about the planet. However, one must remember that our model is greatly simplified, as it is usually assumed that Earth consists of six layers at least, not of only two. On the other hand, it is possible to choose such a core-mass fraction, that the Earth lies exactly on a curve representing two-layer planets made of iron and perovskite silicate when no modification of gravity is introduced (i.e. $\alpha = 0$). We repeated the calculations for different mass fractions and discovered that the best-fit curve is the one for a planet whose iron core contributes only $10\%$ of its total mass (see Fig. \ref{fig3}). This finding is in agreement with the results obtained by Seager et al. (see \cite{seager}).}

{It is also reasonable to ask whether omitting terms containing pressure in the structural TOV equation \eqref{tovJ1} was justified, since the corrections coming from modifications of gravity might be less important that the pressure. In order to check that, we included the pressure in \eqref{tovJ1} and computed the density profiles for a planet of mass equal to one Earth's mass and compared the results. As it turned out, there were no differences exceeding the computational accuracy of our program.}

\section{Conclusions}

In this paper, we have presented numerical solutions to modified Tolman-Oppenheimer-Volkoff equations for non-homogeneous, Earth-like planets in the Palatini $f(\mathcal{R}) = \mathcal{R} + \beta \mathcal{R}^2$ gravity. As it turns out, modifications of the Lagrangian influencing the high-energy limit of the theory, have an impact on non-relativistic regime as well, leading to alteration of interior structure of small, dense objects. This was shown by writing the relevant equations - TOV and mass relation - in terms of conformal invariants, allowing one to analyze the same equations for various theories within one framework. Since the matter is non-relativistic, we neglected pressure in the equations. The last missing piece was the equation of state for the matter, which was used to relate pressure derivative with derivative of the density. We chose a modified polytropic equation of state, following \cite{seager}, for iron and perovskite silicate, which are the main two compounds building the Earth's core and mantle. Numerical integration was performed for a planet whose core mass fraction is 0.3, i.e. after reaching 30$\%$ of total mass, the equation of state was changed from the one for iron to the one for the silicate. As it turned out, the final result is very sensitive to the choice of internal density. A small deviation from the physical density resulted in an incorrect mass-to-radius ratio, therefore some datapoints were excluded from our analysis. That fact did not influence the final result, showing that for different values of the parameters $\beta$, planets of the same mass will have different radii. 

Figures \ref{fig1a} and \ref{fig1b} show that there exists some degeneracy due to the choice of the theory of gravity one is analyzing. The same planet made of certain compounds in the Newtonian theory of gravity might have a different composition in a modified version of the theory, i.e. curves representing different proportions of chemical compounds (or even different substances building the planet in general) in different gravity theories might overlap, so that it might be impossible to say what exactly their properties are and whether they are, for example, capable of supporting life. This effect, while minuscule for small planets, should be significant for bigger ones. For example, for a rocky planet of a mass equal to approximately 37 Earth's masses, which is the case of a Kepler-145b exoplanet \cite{xie}, this effect should have an influence on our prediction of its internal structure, and also on jovian-like planets, as the one discover recently in an external galaxy \cite{stefano}. We also remark that, comparing these new results to the ones obtained in our previous paper \cite{olek2}, we see that due to addition of iron core, the average density of planets increased and now the Earth is significantly above the curve representing the unmodified theory. Now, the best-fit curve would be for $\alpha = 10^{-13}$, but this might be misleading since the realistic model would assume more layers, made of lighter substances.

Figure \ref{fig2} tells us something about the measured structure itself. While we will probably never be able to perform seismic measurements on distant exoplanets, discovering to a bigger accuracy its internal constitution, we have at our disposal some experimental data acquired here on Earth \cite{prem}. We know, to a satisfying accuracy, the internal structure of the Earth, i.e. what layers it is composed of, what their thickness is, chemical composition of each of them, and so on. On the other hand, modifications of gravity influence the thickness of each layer, as it can be seen in Figure \ref{fig2}. Although our analysis was carried out for a much simpler model than Earth, which is usually modelled with at least six different layers, the point is clear: having accurate data obtained from seismic experiments, we can constrain possible values of the parameter $\beta$ (or other parameters for different theories of gravity) by comparing the reference model of Earth to the one predicted by a new gravity theory. One may argue that such corrections introduced by modified gravity will be too small to be detected here on Earth. However this fact can allow to put a constraint on the maximum absolute value of $\beta$, and as the measurements get more accurate, the theory will be confirmed or rejected. 

\vspace{5mm}
\noindent \textbf{Acknowledgement.} 
This work was supported by the EU through the European Regional Development Fund CoE program TK133 "The Dark Side of the Universe". 
AK is a beneficiary of the Dora Plus Program, organized by the University of Tartu.

\end{document}